\documentclass[english]{extarticle}
\usepackage[T1]{fontenc}
\usepackage[latin9]{luainputenc}
\usepackage{amssymb}

\makeatletter
\@ifundefined{date}{}{\date{}}
\makeatother

\usepackage{babel}
\begin{document}
\title{Analysis on the Derivation of the Schrödinger Equation with Analogy
to Electromagnetic Wave Equation}
\author{Xuefeng Bao}
\maketitle

\section{Introduction}

The Schrödinger equation is universally accepted due to its excellent
predictions aligning with observed results within its defined conditions.
Nevertheless, it does not seem to possess the simplicity of fundamental
laws, such as Newton's laws of motion. Various insightful attempts
have been made to elucidate the rationale behind the Schrödinger equation
{[}Derbes 1996, Efthimiades 2006, Field 2010, Ward and Volkmer 2006{]}.
This paper seeks to review existing explanations and propose some
prospectives on the derivation of the Schrödinger equation.

\section{Derivation }

\subsection{General Solution}

Consider a complex wave which propagates in $x$ direction:

\begin{equation}
\psi\left(x,t\right)=Ae^{i\left(kx-\omega t\right)}\label{eq:basic_solution}
\end{equation}
where $\psi$ denotes the wave function, $A$ the amplitude, $k$
the wave number, $\omega$ the angular frequency, $x$ the position,
and $t$ the time. 

We can see that the real part of (\ref{eq:basic_solution}), i.e.,
$Re\left[Ae^{i\left(kx-\omega t\right)}\right]=Acos\left(kx-\omega t\right)$,
is a cosine function that can represent a real wave and is independent
from a wave equation, i.e., we don't need a wave equation to get this
cosine form because this form itself represents a wave that exists
naturally. But, for mathematical convenience, we would prefer to use
its complex version. 

Taking twice partial derivation on $t$ for (\ref{eq:basic_solution})
yields

\begin{equation}
\frac{\partial^{2}}{\partial t^{2}}\psi=-\omega^{2}Ae^{i\left(kx-\omega t\right)}\label{eq:t_2nd_derive}
\end{equation}
and taking twice partial derivation on $x$ we can obtain

\begin{equation}
\frac{\partial^{2}}{\partial x^{2}}\psi=-k^{2}Ae^{i\left(kx-\omega t\right)}.\label{eq:x_2nd_derive}
\end{equation}
Combining (\ref{eq:t_2nd_derive}) and (\ref{eq:x_2nd_derive}), and
considering $k=\frac{2\pi}{\lambda}=\frac{2\pi f}{v}=\frac{\omega}{v}$
(here $\lambda$ is wavelength, $f$ is frequency, and $v$ is velocity),
we have

\begin{equation}
\frac{\partial^{2}}{\partial x^{2}}\psi=\frac{1}{v^{2}}\frac{\partial^{2}}{\partial t^{2}}\psi\label{eq:wave_equation}
\end{equation}
which is a typical wave equation. We can also verify that (\ref{eq:basic_solution})
is solution of (\ref{eq:wave_equation}).

We may notice that (\ref{eq:basic_solution}) is represented by a
pair $\left(k,\omega\right)$, and (\ref{eq:wave_equation}) fits
this representation. What if we change $\left(k,\omega\right)$ to
$\left(E,p\right)$ (here $E$ is energy and $p$ is momentum)? What
wave equation would fit the wave function represented by the new pair
$\left(E,p\right)$? 

\subsection{Electromagnetic Wave}

According to the Photoelectric Effect, each photon in an electromagnetic
wave carries an energy in the form of

\begin{equation}
E=\hbar\omega\label{eq:E_hw}
\end{equation}
and, according to the basic properties of Matter Waves

\begin{equation}
p=\hbar k\label{eq:p_hk}
\end{equation}
and therefore, (\ref{eq:basic_solution}) can be rewritten as 

\begin{equation}
\psi\left(x,t\right)=Ae^{\frac{i}{\hbar}\left(px-Et\right)}.\label{eq:ep_solution}
\end{equation}
Taking $\frac{\partial^{2}}{\partial t^{2}}$ and $\frac{\partial^{2}}{\partial x^{2}}$
on (\ref{eq:ep_solution}), we can get 
\begin{equation}
\frac{\partial^{2}}{\partial t^{2}}\psi=-\frac{E^{2}}{\hbar^{2}}Ae^{i\left(px-Et\right)}\label{eq:e_2nd_derive}
\end{equation}
and 
\begin{equation}
\frac{\partial^{2}}{\partial x^{2}}\psi=-\frac{p^{2}}{\hbar^{2}}Ae^{i\left(px-Et\right)}.\label{eq:p_2nd_derive}
\end{equation}
Relativistic total energy is defined as

\begin{equation}
E^{2}=p^{2}c^{2}+m^{2}c^{4}\label{eq:relative_energy}
\end{equation}
where $c$ denotes the speed of light. As the moving photon does not
carry mass (i.e., $m=0$), (\ref{eq:relative_energy}) can be rewritten
as

\begin{equation}
E^{2}=p^{2}c^{2}.\label{eq:photon_energy_momentum}
\end{equation}
Combine (\ref{eq:e_2nd_derive}), (\ref{eq:p_2nd_derive}) and (\ref{eq:photon_energy_momentum}),
we can get

\begin{equation}
\frac{\partial^{2}}{\partial x^{2}}\psi=\frac{1}{c^{2}}\frac{\partial^{2}}{\partial t^{2}}\psi\label{eq:electromagnetic_wave}
\end{equation}
which is a (plane) wave equation for (monochromatic) electromagnetic
wave.

\subsection{Wave Equation of Particles }

Relativistic total energy of a photon purely equals its kinetic energy,
i.e., (\ref{eq:photon_energy_momentum}). However, this is not true
for a particle with nonzero mass, i.e, (\ref{eq:relative_energy}).
If we mechanically put the momentum and energy of a particle into
(\ref{eq:ep_solution}) and want to establish a wave equation, what
type of energy we should use? The relativistic total energy or the
kinetic energy of the particle? 

\subsubsection{Derivation with relativistic total energy}

Assuming the relationship between momentum and energy is governed
by (\ref{eq:relative_energy}), and substitute it in (\ref{eq:e_2nd_derive}),
we can get

\[
\frac{\partial^{2}}{\partial t^{2}}\psi=-\frac{p^{2}c^{2}}{\hbar^{2}}\psi-\frac{m^{2}c^{4}}{\hbar^{2}}\psi
\]
which can be reformed to

\[
-\frac{p^{2}}{\hbar^{2}}\psi=\frac{1}{c^{2}}\frac{\partial^{2}}{\partial t^{2}}\psi+\frac{m^{2}c^{2}}{\hbar^{2}}\psi
\]
the left hand side exactly equals the right hand side of (\ref{eq:p_2nd_derive}).
Therefore, we have

\begin{equation}
\frac{\partial^{2}}{\partial x^{2}}\psi-\frac{1}{c^{2}}\frac{\partial^{2}}{\partial t^{2}}\psi-\frac{m^{2}c^{2}}{\hbar^{2}}\psi=0\label{eq:KG_equation}
\end{equation}
which is the Klein-Gordon equation.

Then, we may approximate (\ref{eq:relative_energy}) in a way like

\[
E=mc^{2}\left(1+\frac{p^{2}}{m^{2}c^{2}}\right)^{\frac{1}{2}}\approx mc^{2}\left(1+\frac{p^{2}}{2m^{2}c^{2}}\right)=mc^{2}+\frac{p^{2}}{2m}
\]
which is a combination of the rest energy and the classical kinetic
energy. Let's define the classical kinetic energy as:

\begin{equation}
\mathcal{T}=\frac{p^{2}}{2m}.\label{eq:kinetic_energy}
\end{equation}
Therefore, (\ref{eq:ep_solution}) can be written as

\[
\psi=Ae^{\frac{i}{\hbar}\left(px-mc^{2}t-\mathcal{T}t\right)}
\]
or 

\begin{equation}
\psi=e^{\frac{i}{\hbar}-mc^{2}t}\psi_{c}\label{eq:psi_T}
\end{equation}
where

\begin{equation}
\psi_{c}=Ae^{\frac{i}{\hbar}\left(px-\mathcal{T}t\right)}\label{eq:psi_c}
\end{equation}
which is equivalent to (\ref{eq:ep_solution}) when the energy $E$
is purely kinetic energy $\mathcal{T}$. 

Taking $\frac{\partial^{2}}{\partial t^{2}}$ to (\ref{eq:psi_T}),
we obtain

\begin{equation}
\frac{\partial^{2}}{\partial t^{2}}\psi=-e^{-\frac{i}{\hbar}mc^{2}t}\left(\frac{m^{2}c^{4}}{\hbar^{2}}\psi_{c}+\frac{2i}{\hbar}mc^{2}\frac{\partial\psi_{c}}{\partial t}\right)+e^{-\frac{i}{\hbar}mc^{2}t}\frac{\partial^{2}}{\partial t^{2}}\psi_{c}.\label{eq:massive_equation}
\end{equation}
In \cite{ward2006derive}, the following condition seems to be assumed
to hold:

\begin{equation}
\left|-e^{-\frac{i}{\hbar}mc^{2}t}\left(\frac{m^{2}c^{4}}{\hbar^{2}}\psi_{c}+\frac{2i}{\hbar}mc^{2}\frac{\partial\psi_{c}}{\partial t}\right)\right|\gg\left|e^{-\frac{i}{\hbar}mc^{2}t}\frac{\partial^{2}}{\partial t^{2}}\psi_{c}\right|\label{eq:approximate_condition}
\end{equation}
 so that (\ref{eq:massive_equation}) can be approximated as

\begin{equation}
\frac{1}{c^{2}}\frac{\partial^{2}}{\partial t^{2}}\psi=-e^{-\frac{i}{\hbar}mc^{2}t}\frac{m^{2}c^{2}}{\hbar^{2}}\psi_{c}-\frac{2mi}{\hbar}e^{-\frac{i}{\hbar}mc^{2}t}\frac{\partial\psi_{c}}{\partial t}.\label{eq:massive_equation_less}
\end{equation}
Combine (\ref{eq:massive_equation_less}) and (\ref{eq:KG_equation}),
and cancel $\frac{1}{c^{2}}\frac{\partial^{2}}{\partial t^{2}}\psi$,
we can obtain

\[
-e^{-\frac{i}{\hbar}mc^{2}t}\frac{m^{2}c^{2}}{\hbar^{2}}\psi_{c}-\frac{2mi}{\hbar}e^{-\frac{i}{\hbar}mc^{2}t}\frac{\partial\psi_{c}}{\partial t}=\frac{\partial^{2}}{\partial x^{2}}\psi-\frac{m^{2}c^{2}}{\hbar^{2}}\psi
\]
and use (\ref{eq:psi_T}) again, we obtain

\begin{equation}
i\hbar\frac{\partial\psi}{\partial t}=-\frac{\hbar^{2}}{2m}\frac{\partial^{2}\psi}{\partial x^{2}}\label{eq:schrodinger_relativity}
\end{equation}
which is exactly the form of Schrödinger Equation. But, we need to
bear in our mind that the solution of (\ref{eq:schrodinger_relativity}),
i.e., (\ref{eq:ep_solution}) carries relativistic total energy, i.e.,
(\ref{eq:relative_energy}) not $E=\frac{p^{2}}{2m}$. Also, several
approximations are made to get the final conclusion. Dogmatically,
the Klein-Gordon equation is only valid for spin-0 particles. And
the speed of light is a large number but not infinity. Nevertheless,
the derivation above provides a great heuristic inspiration. 

\subsubsection{Derivation with classical kinetic energy}

Assuming that in (\ref{eq:ep_solution}), the energy is purely the
classical kinetic energy, i.e., $E$ in (\ref{eq:ep_solution}) equals
$\mathcal{T}$ in (\ref{eq:kinetic_energy}), or explicitly:

\begin{equation}
E=\frac{p^{2}}{2m}.\label{eq:kinetic_energy_E}
\end{equation}
We can rewrite (\ref{eq:p_2nd_derive}) as

\begin{equation}
-\frac{\hbar^{2}}{2m}\frac{\partial^{2}\psi}{\partial x^{2}}=\frac{p^{2}}{2m}\psi\label{eq:p_2n_derive_clear}
\end{equation}
and $\frac{\partial}{\partial t}$ on (\ref{eq:ep_solution}) and
rearrange we can get

\begin{equation}
-\frac{\hbar}{i}\frac{\partial\psi}{\partial t}=E\psi.\label{eq:E_1st_derive_clear}
\end{equation}
Combining (\ref{eq:p_2n_derive_clear}), (\ref{eq:E_1st_derive_clear}),
and (\ref{eq:kinetic_energy_E}), we have

\[
-\frac{\hbar^{2}}{2m}\frac{\partial^{2}\psi}{\partial x^{2}}=-\frac{\hbar}{i}\frac{\partial\psi}{\partial t}
\]
which is

\begin{equation}
-\frac{\hbar^{2}}{2m}\frac{\partial^{2}\psi}{\partial x^{2}}=i\hbar\frac{\partial\psi}{\partial t}\label{eq:schrodinger_classical}
\end{equation}
which is identical to (\ref{eq:schrodinger_relativity}), but the
$E$ in (\ref{eq:schrodinger_classical}) is the classical kinetic
energy, i,e., $E=\frac{p^{2}}{2m}$. If we want to add the potential
energy ,$V$, into (\ref{eq:schrodinger_classical}), we simply get

\begin{equation}
\left[-\frac{\hbar^{2}}{2m}\frac{\partial^{2}}{\partial x^{2}}+V\right]\psi=i\hbar\frac{\partial\psi}{\partial t}.\label{eq:schrodinger_classical_complete}
\end{equation}
During this derivation, $E=\hbar\omega$ (from (\ref{eq:p_hk}) to
(\ref{eq:ep_solution})), which is for photons, is applied to the
particle with mass. If we assume $E=\frac{1}{2}mv^{2}$, where $v$
is the velocity of the particle, we will get exactly (\ref{eq:wave_equation})
through (\ref{eq:t_2nd_derive}) and (\ref{eq:x_2nd_derive}). But
this is not the Schrödinger equation.

\section{Interpretation}

In the above analysis, we encounter a noteworthy problem. You may
notice that we transformed $\left(k,\omega\right)$ representation
into $\left(E,p\right)$ using (\ref{eq:E_hw}) (photoelectric effect)
and (\ref{eq:p_hk}) (matter waves), and thus, ultimately arriving
at (\ref{eq:ep_solution}). While the matter wave property holds true
for both photons and particles, the same cannot be assumed for the
photoelectric effect. In the above derivations, we directly started
from (\ref{eq:ep_solution}), where (\ref{eq:E_hw}) is implied to
be true. In other words, we imposed the photoelectric effect on particles
by assuming $E=\hbar\omega$ for a particle with mass. Is that a correction
operation? If so, why we can apply the photoelectric effect to particles?
If not, how can we derive the Schrödinger equation, which is widely
considered to be true?

\subsection{Harmonic Oscillation}

When an electron is performing harmonic oscillation, the ground state
energy is $E_{0}=\frac{\hbar\omega_{c}}{2}$, where $\omega_{c}$
is a classical frequency that represents the frequency of the oscillation.
Indeed, it is not precisely $E_{0}=\hbar\omega_{c}$, and it is disturbing
to assume that an eletron is always in a harmonic oscillation. But
at least, we have $E_{0}\propto\hbar\omega_{c}$, Also, we can get
a $\hbar\omega_{c}$-like energy before we assume that he the Schrödinger
equation itself is true. We can start with the Hamiltonian, 

\begin{equation}
\hat{H}=\frac{1}{2m}\hat{p}^{2}+\frac{m\omega_{c}^{2}}{2}\hat{x}^{2}\label{eq:harmonic_hamiltonian}
\end{equation}
which is classical, where $\hat{\cdot}$ denotes operator and $\omega_{c}$
the classical frequency. Applying substitution $\hat{p}\rightarrow\frac{\hbar}{2\bigtriangleup x}$
and $\hat{x}\rightarrow\bigtriangleup x$, where $\bigtriangleup x\bigtriangleup p\geq\frac{\hbar}{2}$
(i.e., Heisenberg uncertainty relation), to (\ref{eq:harmonic_hamiltonian}),
we obtain {[}Bes 2007{]}:

\begin{equation}
E\geq\frac{\hbar^{2}}{8m\left(\bigtriangleup x\right)^{2}}+\frac{m\omega_{c}^{2}\left(\bigtriangleup x\right)^{2}}{2}\label{eq:energy_inequality}
\end{equation}
which means that when $\bigtriangleup x=\left(\frac{\hbar}{2m\omega_{c}}\right)^{\frac{1}{2}}$
we have:
\[
E_{0}=\frac{\hbar\omega_{c}}{2}.
\]

\subsection{Feynman Integration}

If we derive from the Feynman Integration, we still need an assumption
like $\psi=e^{\frac{i}{\hbar}S}$, where $S$ is the action, which
is similar to (\ref{eq:ep_solution}), and this is also the key to
get the Schrödinger equation {[}Field 2010{]}. No one explains why
it can be applied to particles. Also, it's hard to say if Feynman
path proves the Schrödinger equation or it's derived to fit the Schrödinger
equation. Actually, we don't need to be bothered with that, because
the photoelectric effect is more likely to be used to describe a path
(maybe a trajectory of a small part of an electron). Therefore, can
we assume that (\ref{eq:E_hw}) is good for a small part of an electron?
And all the small parts form the electron? It sounds deviant as the
electron is not separable according to the Standard Model. 

\subsection{Virtual Photon}

In fact, a closer examination of the interpretation of the wave function
(\ref{eq:basic_solution}) as a solution to an electromagnetic wave
equation, such as (\ref{eq:electromagnetic_wave}), can offer valuable
insights. Suppose the amplitude of the electromagnetic wave is directly
related to the number of photons propagating in that direction {[}Feynman
1998{]}. In that case, it is conceivable that $\psi$ in (\ref{eq:wave_equation})
is associated with the number of photons, which form a 'beam.' Hence,
when we explore $\psi$ in the context of the Schrödinger Equation
for a particle, it might similarly represent a 'beam,' analogous to
the electromagnetic wave. For the electromagnetic wave, the constituents
of the beam are photons. If we consider a particle to be a 'beam,'
the question arises: what comprises this beam? Are the constituents
also photons? If so, why do we get the Schrödinger equation instead
of an electromagnetic wave equation? A closer examination reveals
that the relationship between energy and momentum for a particle is
defined by $E=\frac{p^{2}}{2m}$ (not $E=pc$), indicating that the
mass of the beam (or each constituent) is non-zero. This is the key
reason that we observe a second-order time derivative in the electromagnetic
wave equation, as opposed to the first-order time derivative in the
Schrödinger equation. The question that naturally follows is: how
should we interpret it?

Virtual photons are widely recognized as a mathematical tool rather
than a physical reality. However, considering electrons as composed
of virtual photons becomes a useful framework for explaining various
issues we have discussed previously. Virtual photons exhibit photon-like
behavior, allowing us to conceptualize an electron as a stream of
virtual photons. Just as the amplitude of an electromagnetic wave
depends on the number of photons, the amplitude in the wave function
of the Schrödinger equation also relies on the number of virtual photons.
Therefore, the $\psi$ in (\ref{eq:electromagnetic_wave}) and (\ref{eq:schrodinger_classical})
share the same interpretation. If electrons interact by exchanging
virtual photons, it may be reasonable to think that virtual photons
are integral components of an electron. Furthermore, virtual photons
can interact with the Higgs field and acquire mass. This could explain
why an electron can be modeled as a beam of virtual light and obeys
$E=\frac{p^{2}}{2m}$. The enigma of why, when measuring an electron,
virtual photons tend to get gathered in locations with the highest
concentration is still mysterious. Nevertheless, this stochastic process
can be understood from a classical perspective {[}Nelson 1966{]},
and the gathering (collapse) process is at the speed of light. Moreover,
it can easily explains the double-slit experiments. Also, photon are
like oscillations according to QFT, and each virtual photon can travel
along a path described by Feynman. 

\section{Conclusion}

In this work, some derivations of the Schrödinger equation were reviewed.
One interesting thing is that in many existing works, $E=\hbar\omega$
is assumed to be true for a particle under the Schrödinger equation.
In fact, if this is true, and combine with the solution for an electromagnetic
wave equation, the Schrödinger equation can be naturally obtained.
Therefore, instead of stating that the Schrödinger equation is correct,
we can rather say that $E=\hbar\omega$ is true for the wave function
of a particle. In this paper, an effort is made to explain how $E=\hbar\omega$
can be true for a particle with mass. 

\section*{References}

$ $

{[}1{]} Ward, D.W. and Volkmer, S.M., 2006. How to derive the schrodinger
equation. arXiv preprint physics/0610121.

{[}2{]} Efthimiades, S., 2006. Physical meaning and derivation of
Schrodinger and Dirac equations. arXiv preprint quant-ph/0607001.

{[}3{]} Derbes, D., 1996. Feynman\textquoteright s derivation of the
Schrödinger equation. American Journal of Physics, 64(7), pp.881-884.

{[}4{]} Field, J.H., 2010. Derivation of the Schrödinger equation
from the Hamilton--Jacobi equation in Feynman's path integral formulation
of quantum mechanics. European journal of physics, 32(1), p.63.

{[}5{]} Bes, D.R., 2007. Quantum mechanics: a modern and concise introductory
course (p. 43). Berlin Heidelberg: Springer.

{[}6{]} Nelson, E., 1966. Derivation of the Schrödinger equation from
Newtonian mechanics. Physical review, 150(4), p.1079.
\end{document}